# Structure Based Aesthetics and Support of Cognitive Tasks for Graph Evaluation

Weidong Huang
University of Tasmania, Australia
Tony.Huang@utas.edu.au

*Abstract*— Drawing principles, or aesthetics, are important in graph drawing. They are used as criteria for algorithm design and for quality evaluation. Current aesthetics are described as visual properties that a drawing is required to have to be visually pleasing. However, most of these aesthetics are originally proposed without consideration of graph structure information. Therefore their ability in visually revealing graph structural features are not guaranteed and indeed mixed results have been reported in the literature regarding their impact on user graph comprehension. In this paper, we propose to derive aesthetics based on graph internal structural features. Further, graphs are often evaluated based on controlled experiments with simple perception tasks to avoid possible confounding factors caused by complex tasks. This leaves their value in supporting complex tasks unevaluated. To fill this gap, we also discuss the possibility of applying evaluation methodologies used in the Cognitive Load Theory research for graph evaluation.

## I. INTRODUCTION

Non-visual graph data are often visualized as node-link diagrams to help users understand them better. However, the same data can be drawn with different layouts. Some layouts can make the diagram look confusing while others do not. As a result, aesthetics criteria have been proposed to define "good" layouts [1]. For example, minimize the number of crossings; distribute nodes evenly; and maximize crossing angles of edges. These aesthetics help make the nodes and edges less cluttered, thus the whole drawing being visually more pleasing. It is commonly accepted that the resulting drawing will be good if a graph is drawn to meet these aesthetic criteria.

However, these aesthetics are originally proposed based on either researchers' personal intuitions or general psychology theories, and they often conflict with each other. Therefore visualizations that are generated to meet these aesthetics are not necessarily always effective for human graph comprehension [2, 22]. Indeed, many graph visualization techniques have been proposed in the literature; it is claimed that those techniques will help users to understand data in one way or another. However, only a few of them have been widely used in practice. More significantly, with the increasingly popular use of graphs in various research fields and everyday life, the need for more effective criteria has become more urgent than ever before.

In response to the need, we have been looking into the criteria we used to evaluate effectiveness of graph drawings and the methods we used to conduct evaluations. It is our observation that:

- Aesthetics that are currently in use are mainly formulated from the perspective of visual properties of node-link diagrams with little consideration of the structure of the graph in question. However, a graph drawing should faithfully reflect and make the graph structural characteristics readily available to viewers to be effective.

- Graphs are mainly evaluated for their ability in assisting users to perform simple perception tasks. However, graph drawings are often an important component of a visualization system to support complex sense making activities. Their ability in supporting those complex activities should also be evaluated.

In the remainder of this paper, we first outline a model of the general graph drawing and evaluation process with a focus on how evaluation is conducted during the process. Then we briefly discuss characteristics of currently used aesthetic criteria and propose to derive new aesthetics based on internal graph structural features. We also briefly review current evaluation methods and propose to evaluate graphs based on how effective a drawing is to support users in performing relatively complex learning and problem solving tasks. The paper concludes with a short summary.

## II. GRAPH DRAWING AND EVALUATION MODEL

When it comes to graph evaluation, the first issue we need to deal with is how to define the quality of graph drawings. It is commonly accepted that data and users are at two sides of a visualization process. On the data side, a quality drawing should visually map the underlying data faithfully for users to read easily. In this regard, aesthetics, usually visual properties, are often used to fulfill this purpose at the early stage of design and no real users are involved in the evaluation. For example, minimum crossings, even distribution of nodes and symmetries. On the end user side, quality can be defined in a range of ways depending on the purpose of the drawing. For example, how long the user will take to perform certain tasks; how much insight the user can take from the drawings, how effective the drawing will be to support users to perform new tasks with new data sets. Desired visual properties are achieved by implementing purposely-designed algorithms, while user–oriented quality is evaluated using corresponding evaluation methods as different quality measures require different evaluation methodologies and processes. A general graph evaluation model can be described in Figure 1.

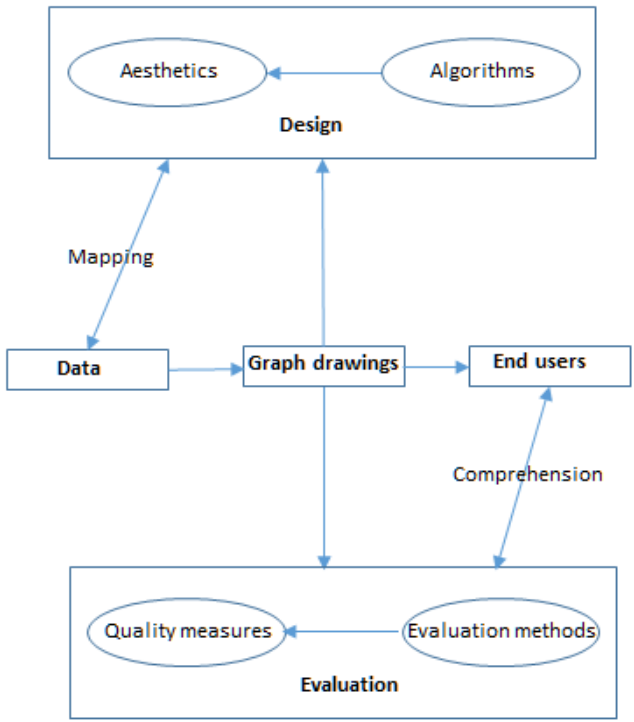

Fig. 1. A graph drawing and evaluation model.

In an attempt to make graph drawing and evaluation process more inclusive and effective and to address the issues we observed and raised in Section 1, we propose to evaluate graphs from a new persepctive. More specifically, from the deisgn perspective, we enrich this model by extending aesthetics to include structural properties and extending quality measures to inculde the support of complex learing and problem solving tasks. From the evaluation perceptive, we propsoe to apply evaluation methods that have been used to evalute instructional methods in the field of Cognitive Load Theory in graph evaluation. More details are provided in the next sections to describe our proposed aesthetic measures and evaluation methods.

## III. AESTHETICS

Empirical studies have shown that intuition-based aesthetics have some limitations in distinguishing differences of effectiveness between drawings; sometimes they can even produce conflicting results [9, 14, 15]. Take the number of edge crossings, one of the most discussed aesthetic criteria, as an example. Figure 2 shows two drawings of a planar graph. The drawing at the top was generated by a planar drawing algorithm and has no crossings, but nodes and edges are squeezed together making paths and the whole drawing difficult to read. On the other hand, the drawing at the bottom was generated by a force-directed algorithm and has some crossings, but despite the crossings, nodes and edges in this drawing are relatively well distributed making relationships of nodes easy to discern. In other words, although it is better in terms of the aesthetic of crossings, the top drawing is not necessarily better than the bottom one in terms of user graph comprehension.

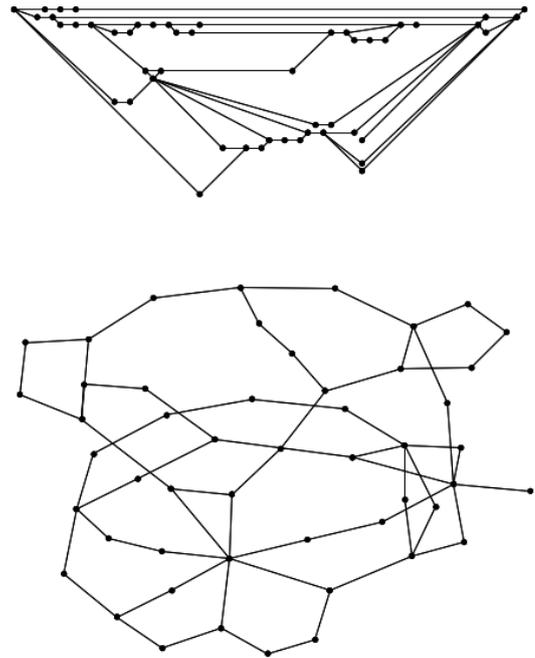

Fig. 2. Two drawings of a planar graph.

A further example is the small evaluation conducted by Mutzel [14]. The author had two drawings of a graph. In these drawings, the crossing patterns differed as the drawing algorithms handled the crossings in different ways during the drawing process. The resulting crossing patterns gave users different impressions on how many crossings the drawing might have and the evaluation showed that indeed this was the case: one drawing with more crossings was perceived having fewer crossings, while the other drawing having fewer crossings was perceived having more crossings. A similar result was also reported by Huang et al. [11] indicating that although crossings are not aesthetically desirable, a group of crossed edges can give a stronger impression to the viewer that relevant nodes belong to the same group. All these examples show that visual layout based aesthetics alone may not be enough to make drawings effective.

Attempts have therefore been made to propose new aesthetics from different perspectives, mainly based on how people read graphs and how people draw graphs. Among them, van Ham and Rogowitz [16] asked users to improve given drawings and found that users often arranged group vertices in a convex hull. Purchase et al. [19] asked users to draw graphs and found that users often aligned vertices and edges to an underlying grid. Yu et al. [17] asked users to draw their personal networks and found that both vertex position and line length were used to differentiate people's positions and relationships. To some extent, these new aesthetics have some connections to graph structures such as group information. To explore along this direction further, we have been looking into deriving aesthetics based on graph structural features. We are

currently conducting two studies, which we briefly describe below:

The first study is on how force-based internal energy is related to task performance. This study makes use of the metaphor of force-directed graph drawing algorithms, which is to treat a graph as a physical system. Although it is generally believed that the smaller energy value a graph drawing has, the better the layout will be, systemic empirical evidence is lacking. We have done a number of user studies in which users were asked to perform a range of graph reading tasks on force-directed drawings of small graphs having up to 100 nodes. Based on the data collected, we are investigating the relationships between measured energy values, combined aesthetics scores and performance. Our preliminary result has indicated that the energy value was negatively correlated with task performance. It should be noted that the overall energy value does not reveal specific structural features, but to some extent, it does allow us to define how a structural feature is to be visualized if acting forces are properly defined.

The second study is to investigate possible benefits of drawing graphs with Euclidean distances of nodes reflecting their theoretic distances (length of the shortest path). In other words, we expect that drawings are better if graphs are drawn with the theoretic distance of nodes being proportional to their Euclidean distance. This is mainly based on the fact that most of graph reading tasks involve path tracing and that viewers tend to follow paths that align with the geometric path of nodes [2]. We call this drawing principle as a Distance aesthetic criterion. This aesthetic implies a commonly referred drawing requirement that nodes of a group should be positioned close to each other. The Distance aesthetic can be measured as either the overall graph stress [18] or a local ratio of the Euclidean distance and theoretic distance of relevant nodes. It should be noted that drawing graphs with low stress is a researched topic and it has been observed that drawings with lower stress have better layout quality (e.g., [18]). Again, this observation is not backed by user studies and how it is related to human graph comprehension is not entirely clear. An exception is the recent study reported by Chimani et al. [20]. In their study, user preference is evaluated with drawings of relatively large graphs with up to one thousand nodes. A negative correlation between stress and preference was reported: people preferred low stress drawings more than those with high stress.

## IV. EVALUATION

It has been acknowledged that evaluation is an important part of a visualization process [3]. In evaluating graph aesthetics, there are mainly two types of research which include qualitative user preference studies based on questionnaires and quantitative task performance experiments based on controlled or semi-controlled design. Among them, controlled within-subject design is the most frequently used for the purpose of validation of aesthetics. For this type of design, same graphs are drawn a few times to change the value of the aesthetic in consideration, and at the same time to keep other known aesthetics or visual factors unchanged. Although this approach has been successful in validating aesthetics, in theory, keeping other aesthetics unchanged is difficult to achieve. And this is particularly true in the case of graphs. This is because nodes are linked to each other; changing the score of one aesthetic will unavoidably change scores of others.

Another limitation is that to make experiments controllable, simple fact finding tasks are often used in the experiment design. This is due to the consideration of likely interference of confounding factors. However, the purpose of visualization goes beyond finding factors about the data in the diagram through simple perceptual or cognitive tasks. Real tasks with visualizations are often complex and difficult requiring intensive and prolonged information processing in human memory, such as learning, problem solving and decision making. These types of tasks are usually time-consuming, taking time longer than just a few seconds or minutes. Therefore, visualizations should also be evaluated for their effectiveness in facilitating performance of such complex tasks. The lack of complex tasks in evaluation has been a great threat to the generality of evaluation findings.

As pointed out by Chang et al. [5], "while many potentially useful methodologies have been proposed, there remain significant gaps in assessing the value of the open-ended exploration and complex task-solving that the visualization community holds up as an ideal." To deal with the above-mentioned limitations, different evaluation methods should be employed. In fact, a number of innovative evaluation methods and scenarios have been proposed in order to better capture the value of visualizations [3, 4]. Among them, Ware et al. [9] introduced a method that can be used to evaluate different optimization criteria without having to manipulate scores of the criteria. Shneiderman and Plaisant [8] described MILCs that allow researchers to evaluate visualizations in more naturalistic and creative situations for a longer period time so that their support for experts to deal with difficult problems can be better evaluated and understood. Saraiva et al. [21] proposed an insight-based approach that captures the entire analysis process of users using visualization tools to seek insights into the data during a longitudinal study. Chang et al. [5] proposed a learning-based evaluation framework that evaluates how visualizations can help users to solve new tasks.

Recently, there have been some efforts in evaluating visualizations using cognitive measures (e.g., [2, 6, 7]). Particularly, we have introduced cognitive load and visualization efficiency measures inspired by the Cognitive Load Theory (CLT) research in Educational Psychology [10]. We also noticed that methods used in empirical research on CLT could provide a solution to the issue of lacking complex tasks in graph evaluation. CLT is built on a theoretical framework of human cognitive architecture that includes a limited working memory for information processing and an unlimited long-term memory for information storage [13]. It considers cognitive load associated with learning as a major factor determining the success of instructional methods and is one of the fundamental theories that are used to describe the cognitive processes in complex learning activities in order to derive design principles for effective instruction.

Empirical studies on CLT are typically conducted in learning environments. Students are first handed out learning materials in which a piece of new knowledge has been integrated into two or more instructional formats for the purpose of comparison. Then students in different groups are instructed to go through these materials in either natural or simulated learning settings. During the process, their performance and mental effort data are collected for final data processing and analysis. One of the pivotal assumptions for the experimental design is that the format of learning materials has a direct causal relationship with cognitive consequences of learning. The method commonly adopted in CLT research is described as follows:

1. Propose research questions and formulate hypotheses based on well-accepted design principles in relation to the assumed human cognitive architecture.

2. Conduct tutorials of the same learning materials presented with different instructional methods.

3. Obtain mental effort ratings after the completion of the learning phase.

4. Hand out test questionnaire that consists of different types test items. These items are usually categorized into recall, near-transfer and far-transfer ones.

5. Obtain mental effort ratings either after each question or after the completion of the whole test phase.

6. Measure performance including time and scores and compute instruction efficiency based on the collected performance and mental effort data.

7. Conduct statistical analysis, discuss the results and draw conclusions.

The experiments can be between-subject or within-subject. The whole experiment design is based on the analysis of how tasks will be executed in relation to the format of instruction and its implications on the limited working memory. The implications are measured in terms of mental effort, task performance and instruction efficiency. Regarding the method described above, learning tasks are complex and instructional conditions are not necessarily rigorously controlled as pointed out by van Merrienboer and Sweller [12]. However, more than two decades of extensive research in this area has demonstrated that the experimental design based on human cognitive architecture and cognitive load measures does yield valid and important principles for instructional design.

We believe similar approaches can be used for graph evaluation by focusing the design of experiments around the human cognitive architecture. It should be noted that this CLT based approach is similar but different to the "learning-based evaluation methodology" of Chang et al. [5]. These two approaches are similar in that both are set out to evaluate the value of visualization for supporting exploration and problem solving tasks, while they are different in that the former is still performance based while the latter does not directly measure task performance.

## V. CONCLUDING REMARKS

Empirical graph drawing research has gained increasing attention in recent years. On the one hand, we see more and more user studies being conducted for newly proposed visualization techniques. On the other hand, there is a need for new evaluation methodologies or frameworks to accommodate different needs of studies so that the value of visualization can be fully captured and evaluated in an appropriate manner. In this paper, we discussed two important aspects of graph evaluation: evaluation criteria (aesthetics) and evaluation methods. For aesthetics, we believe that aesthetics should be evaluated based on not only how visually pleasing the resulting drawings are, but also how well the drawings can visually convey graph structural information. More specifically, we proposed to use the commonly understood concepts of energy and stress of graphs as aesthetic criteria of overall diagram layout. Further, the Distance aesthetic is suggested based on a drawing practice of laying out nodes in such a way that the Euclidean distance of any two nodes is proportional to their theoretic distance. Generally speaking, perhaps the real research question regarding aesthetics is to find the right visual properties that faithfully reflect the underlying structure features.

For the methodology of evaluation, we proposed to evaluate visualizations for their support of exploration and problem solving using the theoretic framework of human cognitive architecture established in CLT. It is commonly accepted that the value of visualization lies beyond supporting fact-finding or simple perceptual tasks. Successful visualizations should also support users to explore further information that is not readily available and make correct decisions based on the information collected. When complex cognitive tasks such as these are involved, limitations of working memory in processing information stands out and should be taken into consideration. As a result, it is natural to use cognition based methods for evaluation if complex tasks are to be used as part of evaluation. This will also help maximize the external validity of the study and improve the generality of experimental findings. We hope that visualization researchers will find the cognition based experimental design and cognitive load based measurements useful and practical. It should allow researchers to conduct more realistic experiments and assess the effectiveness of visualizations more precisely.